\documentclass[conference, a4paper]{IEEEtran}

\makeatletter
\def\@@@@IEEEcomsocverifymathfont{%
	\typeout{-- Forcing newtxmath instead of mathtime.}%
	\RequirePackage{newtxtext,newtxmath}%
}
\makeatother
\ifCLASSINFOpdf
   \usepackage[pdftex]{graphicx}
   \graphicspath{{../pdf/}{../jpeg/}}
   \DeclareGraphicsExtensions{.pdf,.jpeg,.png}
\else
   \usepackage[dvips]{graphicx}
   \graphicspath{{../eps/}}
   \DeclareGraphicsExtensions{.eps}
\fi
\usepackage{nopageno}   
\usepackage{fixltx2e}
\usepackage{amsfonts}
\usepackage{amsmath,amssymb,amsfonts}
\usepackage{cite} 
\usepackage{subfigure} 
\usepackage{amsmath,amssymb,amsfonts}
\usepackage{algorithmic}
\usepackage{graphicx}
\usepackage{textcomp}
\usepackage{algorithm}
\usepackage{multirow}
\usepackage{algorithmic}
\usepackage{bm}
\usepackage{cite}
\usepackage{float}
\usepackage{url}
\usepackage[pdftex]{graphicx}
\usepackage{diagbox}
\usepackage{graphics}
\usepackage{setspace}
\usepackage{caption}
\usepackage{threeparttable}
\usepackage{textcomp}
\usepackage{fancyhdr}

\usepackage[T1]{fontenc}
\usepackage{aecompl}
\IEEEoverridecommandlockouts
\begin{document}
%

\title{A Secure Affine Frequency Division Multiplexing for Wireless Communication Systems}





\author{
		\IEEEauthorblockN{Ping Wang, Zulin Wang, Yuanfang Ma, Xiaosi Tian and Yuanhan Ni\thanks{* Corresponding author.}\IEEEauthorrefmark{1} }
		\IEEEauthorblockA{School of Electronic and Information Engineering, Beihang University, Beijing, 100191, China\\
				Email:wangping\_119@buaa.edu.cn, wzulin@buaa.edu.cn, yuanfangma@buaa.edu.cn,  \\
				 xiaosi\_tian@buaa.edu.cn,  yuanhanni@buaa.edu.cn \\
		}
	}

\maketitle

\thispagestyle{fancy}
\fancyhead{}
\lhead{}
\cfoot{}
\rfoot{}

\begin{abstract}
This paper introduces a secure affine frequency division multiplexing (SE-AFDM) for wireless communication systems to enhance communication security. Besides configuring the parameter $c_1$ to obtain communication reliability under doubly selective channels, we also utilize the time-varying parameter $c_2$ to improve the security of the communications system. The derived input-output relation shows that the legitimate receiver can eliminate the nonlinear impact introduced by the time-varying $c_2$ without losing the bit error rate (BER) performance. Moreover, it is theoretically proved that the eavesdropper cannot separate the time-varying $c_2$ and random information symbols, such that the BER performance of the eavesdropper is severely deteriorated. Meanwhile, the analysis of the effective signal-to-interference-plus-noise ratio (SINR) of the eavesdropper illustrates that the SINR decreases as the value range of $c_2$ expands. Numerical results verify that the proposed SE-AFDM waveform has significant security while maintaining good BER performance in high-mobility scenarios.


\end{abstract}



%
\IEEEpeerreviewmaketitle

\section{Introduction}
Sixth-generation (6G) wireless communication networks are expected to provide plentiful usage scenarios, enabling advanced capabilities such as wide area coverage, low latency of 0.1 ms, peak throughput of 1 Tb/s, etc. With ubiquitous connectivity, extended coverage and increased terminals expose more sensitive information to eavesdropping risks in open wireless communication environments. Furthermore, low latency and high peak throughput in Hyper Reliable and Low-Latency Communications
(HRLLC) constrain the application of security strategies with high latency and high complexity \cite{ara2024physical}. Therefore, the security of 6G has attracted extensive research.

The well-known strategy for wireless communication security is to implement encryption in the network or application layer, which has been widely used in the military, medicine, and other fields. However, key distribution is a challenging task in decentralized and heterogeneous networks, and the high computational complexity of encryption and decryption may lead to extra latency and limited throughput \cite{solaija2022towards}. Thus, it is an open question for implementing encryption to meet the requirements of low latency and peak throughput in 6G, especially in networks with limited computational capabilities and constrained terminal sizes \cite{liu2022ensuring}.

Besides the network layer and application layer, the physical layer (PHY) can provide wireless communication security, i.e., physical layer security (PLS) \cite{wyner1975wire}. With higher scalability and lower complexity, PLS techniques have caught significant research attention. Due to the advantage of not requiring additional power, ingeniously designed secure waveforms have been widely studied to strengthen PLS security. In the Global Positioning System (GPS), a spread spectrum-based secure waveform based on long-period Pseudo-noise (LPPN) sequences can ensure satellite communication security. The fact of ensuring security is eavesdroppers are unable to synchronize with the LPPN sequence, while the legitimate receiver can \cite{raju2012digital}. However, the spectrum efficiency of the spread spectrum-based waveform may be decreased. Since orthogonal frequency division multiplexing (OFDM) can achieve high spectrum efficiency, an OFDM-based secure waveform is presented to enhance wireless security by modifying the subcarrier spacing for each information symbol, namely improved spectrum efficient frequency division multiplexing (SEFDM) \cite{xu2021waveform}. However, the bit error rate (BER) performance of the OFDM-based waveform deteriorates in doubly selective channels \cite{yuan2024papr}.

Due to the advantage of orthogonal time frequency space (OTFS) in obtaining full diversity over doubly selective channels \cite{OTFShadani2017orthogonal}, various OTFS-base secure waveforms have been researched \cite{sun2021orthogonal,sun2021secure}. Specifically, by spreading the information symbols in either the delay or Doppler domain, a secure OTFS-based waveform, namely DS-OTFS, is formed to improve communication security at the expense of diminished spectrum efficiency \cite{sun2021orthogonal}. Moreover, by rotating information symbols in the delay-Doppler domain based on the legitimate channel, R-OTFS is proposed to achieve secure communication by leveraging the channel diversity to reduce the signal-to-interference-plus-noise ratio (SINR) for the eavesdropper \cite{sun2021secure}. However, in R-OTFS, the strong correlation between the legitimate and the eavesdropping channels may undermine the security of the communication system. Furthermore, due to the excessive pilot overhead caused by the two-dimensional (2D) structure of OTFS, OTFS-based secure waveforms improve communication security at the cost of reduced spectrum efficiency.





%


Affine frequency division multiplexing (AFDM) with one-dimensional (1D) pilots is proposed, which can achieve higher spectrum efficiency and the same BER performance compared to OTFS \cite{bemani2023affine}. By adjusting two discrete affine Fourier transform (DAFT) parameters, i.e., $c_1$ and $c_2$, AFDM can be fully compatible with OFDM, making AFDM regarded as one of the candidate waveforms for 6G \cite{zhou2024overview}. AFDM-based researches mainly focus on improving the reliability of communications by tuning $c_1$, such as channel estimation \cite{yin2022pilot}, equalization \cite{bemani2022low}, integrated sensing and communications \cite{ni2022afdm}, etc. Recently, the parameter $c_2$ of AFDM is adjusted to reduce peak-to-average power ratio (PAPR) \cite{yuan2024papr} or improve spectrum efficiency \cite{zhu2023design}. However, the security of AFDM waveforms is insufficient to meet the demands of 6G.

In this paper, we propose a secure affine frequency division multiplexing (SE-AFDM) with time-varying $c_2$ to enhance security while maintaining the reliability of communications systems. 
In SE-AFDM system, the time-varying $c_2$ is generated from a codebook with a value range according to an index controlled by a LPPN sequence. The LPPN sequence can be reconstructed by the legitimate receiver to synchronize the time-varying $c_2$, while the eavesdropper cannot generate synchronized $c_2$ due to the unknown of the LPPN sequence. The theoretical derivation confirms that the impact of the time-varying $c_2$ can be eliminated at the legitimate receiver with the synchronized $c_2$, but the eavesdropper cannot separate the time-varying $c_2$ and random information symbols. Furthermore, the analysis of the effective SINR of the eavesdropper reveals that expanding the value range of $c_2$ can decrease the effective SINR of the eavesdropper. The simulation results demonstrate the BER performance of the eavesdropper approaches 0.5 with an appropriately designed value range of $c_2$, while the legitimate receiver achieves the same BER performance and spectrum efficiency compared to the existing AFDM system.

\section{Preliminaries}


\subsection{Basic Concepts of AFDM}

Firstly, the basic concepts of AFDM proposed in \cite{bemani2023affine} are briefly reviewed. We use $\mathbf{x}$ to denote an $N {\times} 1$ vector of quadrature amplitude modulation (QAM) symbols. 
The $N$ points inverse discrete affine Fourier transform (IDAFT) is performed to map $\mathbf{x}$ from the affine Fourier transform (AFT) domain to the time domain, i.e.,\cite{bemani2023affine}
\begin{equation}
	{s}\left[{n}\right] = \frac{1}{{\sqrt N }}\sum\limits_{m = 0}^{N - 1} {{x}\left[ {m} \right]} {e^{j2\pi \left( {{c_1}{n^2} + \frac{mn}{N} + {c_2}{m^2}} \right)}} ,
\end{equation}
where $c_1$ and $c_2$ are the AFDM parameters, and $n=0,\ldots,N {-} 1$. Then, a chirp-periodic prefix (CPP) with a length of $N_{cp}$ is added,
which is defined as\cite{bemani2023affine}
\begin{equation}\label{eq:symbol_AFDM}
	{s}\left[ {n} \right] {=} {s}\left[ {n {+} N} \right]{e^{ - i2\pi {c_1}\left( {{N^2} + 2Nn} \right)}},n =  - {N_{cp}}, \ldots , - 1.
\end{equation}


Then, AFDM signal is transmitted over a communication channel with $P$ paths, in which the gain coefficient, time delay and Doppler shift of the $i$-th path are denoted by ${h_i}$, ${\tau _i}$, ${f_{d,i}}$, respectively. The received signal vector in the time domain is given by \cite[Eq. (6)]{wu2022integrating}
\begin{equation}\label{eq:received_sig_time}
	{r}\left[ n \right] = \sum\limits_{i = 1}^P {{{\tilde h}_i}} {s}\left[ {n - {l_i}} \right]{e^{j2\pi {f_i}n}} + {{w}}_t\left[ n \right],
\end{equation}
where $\mathbf{{w}}_t \hspace{-0.5ex}\sim \hspace{-0.5ex}\mathcal {CN}\left( {0, \sigma_{c} ^2\mathbf{I}} \right)$ is an additive Gaussian noise vector, ${{\tilde h}_i} {=} {h_i}{e^{ - j2\pi {f_{d,i}}{\tau _i}}}$, ${l_i} {=} {{{\tau _i}} \mathord{\left/
		{\vphantom {{{\tau _i}} {{t_s}}}} \right.
		\kern-\nulldelimiterspace} {{t_s}}}$, ${f_i} {=} {f_{d,i}}{t_s}$ with ${t_s}$ denoting the sampling interval, and $n \hspace{-0.5ex}\in \hspace{-0.5ex}\left[ { - {N_{cp}},N - 1 } \right]$.

After discarding CPP and performing $N$ points discrete affine Fourier transform (DAFT), the resulted signal in the AFT domain can be written as\cite{bemani2023affine}
\begin{equation}
	\mathbf{y} = {\mathbf{H}_{\rm eff}}\mathbf{x} + \mathbf{{w}}_a= \sum\limits_{i = 1}^P {{\tilde h_i}{\mathbf{H}_{A,i}}\mathbf{x}} + \mathbf{{ w}}_a,
\end{equation}
where ${\mathbf{H}_{\rm eff}} {=} \mathbf{A}{\mathbf{H}_{c,t}}\mathbf{A}^{{H}}$ with ${\mathbf{H}_{c,t}} = \sum\limits_{i = 1}^P {{\tilde h_i}{\bm{\Gamma} _i}{\bm{\Delta} _{{f_i}}}{\bm{\Pi} ^{\left({l_i}\right)}}}$ being the communication channel matrix in the time domain, $\bm{\Pi}$ is the forward cyclic-shift matrix, ${\bm{\Gamma} _i} = {\rm diag}\left( {\left\{ {\begin{array}{*{20}{c}}
			{{e^{ - i2\pi {c_1}\left( {{N^2} - 2N\left( {{l_i} - n} \right)} \right)}}},&{n < {l_i}},\\
			1,&{n \ge {l_i}}.
	\end{array}} \right.} \right)$, ${\bm{\Delta} _{{f_i}}} = {\rm diag}\left( {{e^{ i2\pi {f_i}n}},n \in \left[0, {N-1}\right]} \right)$, ${\mathbf{H}_{A,i}}=\mathbf{A}{\bm{\Gamma} _i}{\bm{\Delta} _{{f_i}}}{\bm{\Pi} ^{\left({l_i}\right)}}\mathbf{A}^{{H}}$, $\mathbf{A} = {\bm{\Lambda} _{{c_2}}}\mathbf{F}{\bm{\Lambda} _{{c_1}}}$,  $\mathbf{F}$ is the discrete Fourier transform (DFT) matrix,\hspace{-0.5ex} ${\bm{\Lambda} _{{c_i}}} {=}  {\rm diag} {\left( {{e^{ - j2\pi c_i{n^2}}},n = 0, \ldots ,N {-} 1}, i=1,2 \right)}$,  and $\mathbf{{w}}_a=\mathbf{A}\mathbf{w}_t$. ${{H}_{A,i}}\left[ {p,q} \right]$ is given by\cite{bemani2023affine}
\begin{eqnarray}
	{{H}_{A,i}}\left[ {p,q} \right] = \frac{1}{N}{e^{j\frac{{2\pi }}{N}\left( {N{c_1}l_i^2 - ql_i + N{c_2}\left( {{q^2} - {p^2}} \right)} \right)}}{\mathcal{F}}_i\left[ {p,q} \right],
\end{eqnarray}
where ${{\mathcal{F}}_i}\left[ {p,q} \right] {=} \frac{{{e^{ - j2\pi \left( {p - q - {\nu_i} + 2N{c_1}{l_i}} \right)}} - 1}}{{{e^{ - j\frac{{2\pi }}{N}\left( {p - q - {\nu_i} + 2N{c_1}{l_i}} \right)}} - 1}}$ with ${\nu _i} = N{f_i} = \frac{{{f_{d,i}}}}{{\Delta f}} = {\alpha _i} + {a_i} \in \left[ { - {\nu _{\max }}, {\nu _{\max }}} \right]$ being the Doppler shift normalized with respect to the subcarrier spacing ${\Delta f}$, and ${\alpha _i} {\in} \left[ { - {\alpha _{\max }},{\alpha _{\max }}} \right]$ and ${a_i} {\in} \left( { - \frac{1}{2},\frac{1}{2}} \right]$ represent the integral and fractional part of ${\nu _i}$, respectively. $p$ and $q \in \left[ 0,N-1 \right]$, 

For the integral normalized Doppler shift case, i.e., $a_i{=}0$, there is only one non-zero element in each row of $\mathbf{H}_{A,i}$, i.e., 
\begin{equation}
	{{H}_{A,i}}\left[ {p,q} \right] = \left\{ {\begin{array}{*{20}{c}}
			\hspace{-0.2cm}{{e^{j\frac{{2\pi }}{N}\left( {N{c_1}l_i^2 - ql_i + N{c_2}\left( {{q^2} - {p^2}} \right)} \right)}}},\hspace{-1.5ex}&{q = {{\left\langle {p + lo{c_i}} \right\rangle}_N}},\\
			0,&{otherwise},
	\end{array}} \right. 
\end{equation}
where $lo{c_i} {=} {{{\left\langle {2N{c_1}{l_i} {-} {\alpha _i}} \right\rangle}_N}}$. Hence, the input-output relation in the AFT domain is given by 
\begin{align}\label{eq:in_out_relation_2}
	&{y}\left[ {p} \right] \hspace{-0.5ex} = \hspace{-0.5ex}\sum\limits_{i = 1}^P {{\tilde h}_i}
	{e^{j\frac{{2\pi }}{N}\left( N{c_1}l_i^2 {- ql_i + N{c_2}\left( {{{q}^2} - {p^2}} \right)} \right)}}{x}\left[ {q} \right] \hspace{-0.5ex}{+} { w}_a\left[ {p} \right],
\end{align}
where ${q {=} {{\left\langle {p + lo{c_i}} \right\rangle}_N}}$ and $p\in \left[0,{N-1}\right]$.
For the fractional normalized Doppler shift case, 
there are some non-zero elements and the peak is still at ${q {=} {{\left\langle {p + lo{c_i}} \right\rangle}_N}}$ in each row of $\mathbf{H}_{A,i}$\cite{bemani2023affine}.

\section{Secure Affine Frequency Division Multiplexing}

In this section, we introduce an SE-AFDM system to improve the security of communication systems.
The input-output relation of SE-AFDM at the Bob is derived. 

\subsection{Proposed SE-AFDM System}

In this paper, we use Alice, Bob and Eve to denote the transmitter, the legitimate receiver and the eavesdropper, respectively.




\vspace{0.5ex}
\noindent\emph{(1) Modulation at the Alice}
\vspace{0.5ex}

\begin{figure*}[htbp]
	\centering	
	\includegraphics[width=6.0in]{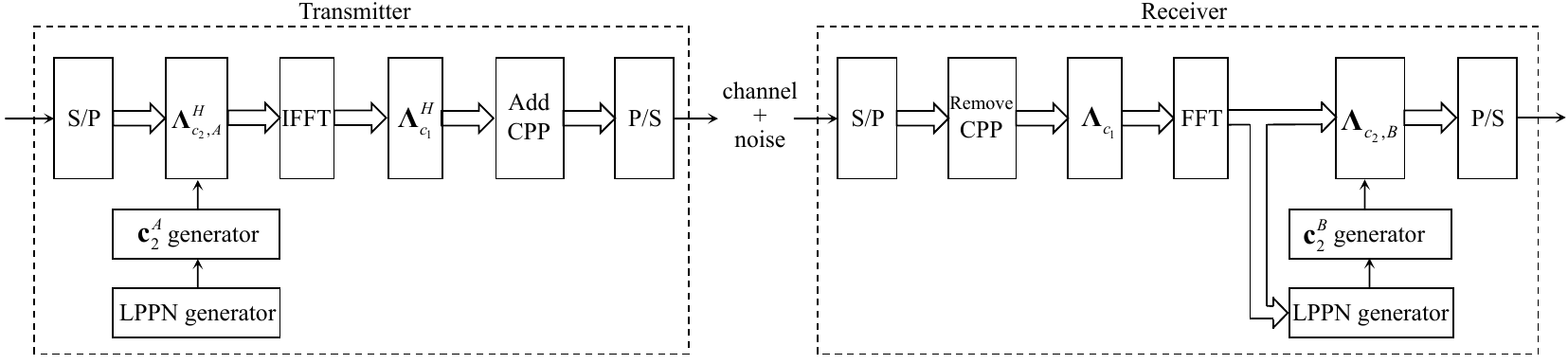}
	\caption{Block diagram of SE-AFDM communication system. 
		\label{fg:SE_AFDM}}
	\vspace{-10pt}
\end{figure*}
The corresponding block diagram of the SE-AFDM communication system is shown in Fig. \ref{fg:SE_AFDM}.
Consider an $N\times 1$ information symbol vector ${x}\left[ n \right]$, $n=0,1,\ldots ,N-1$ from a modulation alphabet $\mathbb{A}=\left\{ {{a}_{1}},\ldots ,{{a}_{\left| \mathbb{A} \right|}} \right\}$ (e.g. QAM), which are arranged on the affine domain. Firstly, the vector $\mathbf{x}$ is multiply by a matrix ${\bm{\Lambda} ^{H}_{{c_2,A}}} {=} {\rm diag}\left( {{e^{ - j2\pi {c}^A_2\left[m\right]{m^2}}},m = 0, \ldots ,N {-} 1} \right)$, where $\mathbf{c}^A_2$ is an ${N \times 1}$ parameter vector at the Alice, and ${c}^A_2\left[m\right]$ denotes the parameter $c_2$ corresponding the $m$-th subcarrier. Each element of $\mathbf{c}^A_2$ is chosen from a codebook $\mathbb{C}_2 = \left\{c_{2,1},c_{2,2},\cdots,c_{2,M} \right\}$ according to an index that is generated by converting a ${log}_{2}{M}$-bit binary sequence to decimal, where $M$ denotes the size of $\mathbb{C}_2$. The ${log}_{2}{M}$-bit binary sequence is obtained by sequentially truncating a LPPN sequence that is produced by the LPPN generator at the Alice. And the codebook $\mathbb{C}_2$ is pre-designed by uniformly discretizing the value range of $c_2$ within [$-c_{2,\rm max}$, $c_{2,\rm max}$] which is public to everyone (including Alice, Bob and even Eve). 


And then, subsequent operations are the same as the existing AFDM waveform, i.e., performing IDFT, multiplying by matrix ${\bm{\Lambda} ^{H}_{{c_1}}}$, adding CPP. The resulting SE-AFDM waveform at the Alice in the time domain can be written as
\vspace{-0.5ex}
\begin{equation} \label{eq:ES_AFDM_Tx}
	{s}_{A}\left[ n \right]=\frac{1}{\sqrt{N}}\sum\limits_{m=0}^{N-1}{{x}}\left[ m \right]{{e}^{j2\pi \left( {{c}_{1}}{{n}^{2}}+{c}^A_2\left[m\right]{{m}^{2}}+\frac{mn}{N} \right)}} ,
\end{equation}
\vspace{-0.5ex}
where $n = - {N_{cp}}, \ldots , N- 1$. 

\vspace{0.5ex}
\noindent\emph{(2) Demodulation at the Bob}
\vspace{0.5ex}

After transmission over the channel with $P$ paths whose gain
coefficient, time delay and Doppler shift of the $i$-th path are denoted by ${h_i^B}$, ${\tau _i^B}$, ${f_{d,i}^B}$, respectively, the received signal vector at the Bob in the time domain is given by
\vspace{-1ex}
\begin{equation}\label{eq:received_sig_Bob_time}
{r}_B\left[ n \right] = \sum\limits_{i = 1}^P {{{\tilde h}^B_i}} {s}_A\left[ {n - {l_i^B}} \right]{e^{j2\pi {f_i^B}n}} + {{w}}_B\left[ n \right],
\end{equation}
where ${{{\tilde h}^B_i}} = {h_i^B}{e^{ - j2\pi {f^B_{d,i}}{\tau ^B_i}}}$,  ${l^B_i} {=} {{{\tau^B_i}} \mathord{\left/{\vphantom {{{\tau ^B_i}} {{t_s}}}} \right.
\kern-\nulldelimiterspace} {{t_s}}}$, ${f^B_i} {=} {f^B_{d,i}}{t_s}$ with ${t_s}$ denoting the sampling interval, and $\mathbf{w}_{B}\in {{\mathbb{C}}^{N\times 1}}$ is an additive Gaussian noise vector with power spectral density $\sigma _{n,B}^2$.

After serial to parallel conversion (S/P) and discarding CPP, the received SE-AFDM signal at the Bob in the time domain is given by 

\begin{equation}
{\mathbf{r}}_{B} = \sum\limits_{i = 1}^P {{{\tilde h}_i^B}{{\bf{\Gamma }}_{{\rm cpp}{_i}}}{{\bf{\Delta }}_{{f_{i}^B}}}{{\bf{\Pi }}^{{l_i^B}}}{\bf{s}}_{A}}  + {\bf{w}}_{B}.
	\label{eq: relation between R and x in matrix form} 
\end{equation}

Then, multiplying by matrix ${\bm{\Lambda} _{{c_1}}}$ and performing DFT, we can get
\vspace{-0.2cm}
\begin{eqnarray}   
\begin{aligned} \label{eq:ES_AFDM_Rx_TD}
\begin{split}
	&{\mathbf{r}}'_{B}\hspace{-1mm}=\hspace{-1mm} \sum\limits_{i = 1}^P {{{\tilde h}^B_i} \mathbf{F}{\bm{\Lambda} _{{c_1}}} {{\bf{\Gamma }}_{{\rm cpp}{_i}}}{{\bf{\Delta }}_{{f^B_{i}}}}{{\mathbf{\Pi }}^{{l^B_i}}}{\bm{\Lambda} ^{H}_{{c_1}}} \mathbf{F}^{ H}{\bm{\Lambda} ^{H}_{{c_2,A}}}\mathbf{x} }  + {\mathbf{w}}'_{B},\hspace{-1mm}
\end{split}
\end{aligned}
\end{eqnarray}
where ${\bf{w}}'_{B} = \mathbf{F}{\bm{\Lambda} _{{c_1}}} {\bf{w}}_{B}$. 

After that, ${\mathbf{r}}'_{B}$ is multiplied by the matrix ${\bm{\Lambda} _{{c_2,B}}} {=}  {\rm diag}\left( {{e^{ - j2\pi {c}^B_2\left[m\right]{m^2}}},m = 0, \ldots ,N {-} 1} \right)$, where $\mathbf{c}^B_2$ is an ${N \times 1}$ parameter vector at the Bob.
Each element of $\mathbf{c}^B_2$ is also chosen from the codebook $\mathbb{C}_2$ according to an index that is controlled by a ${log}_{2}{M}$-bit random sequence. The random sequence is produced by another LPPN generator at the Bob.
Now, the received matrix at the Bob in the affine domain can be written in matrix form as
\vspace{-0.5ex}
\begin{align} \label{eq:ES_AFDM_Rx_AD}
	\mathbf{y}_{B}&= \sum\limits_{i = 1}^P {{{\tilde h}^B_i}{\bm{\Lambda} _{{c_2,B}}} \mathbf{F}{\bm{\Lambda} _{{c_1}}} {{\bf{\Gamma }}_{{\rm cpp}{_i}}}{{\bf{\Delta }}_{{f^B_{i}}}}{{\bf{\Pi }}^{{l^B_i}}}{\bm{\Lambda} ^{H}_{{c_1}}} \mathbf{F}^{H}{\bm{\Lambda} ^{ H}_{{c_2,A}}}\mathbf{x} }  + {\mathbf{\bar w}}_{B} \nonumber \\
	& = \mathbf{H}_{{\rm eff},B} \mathbf{x}  + {\bf{\bar w}}_{B},
\end{align} 
where ${\bf{\bar w}}_{B} = {\bm{\Lambda} _{{c_2,B}}}\mathbf{F}{\bm{\Lambda} _{{c_1}}} {\bf{w}}_{B}$. 

\vspace{-1ex}

\subsection{Input-Output Relation of SE-AFDM Between Alice and Bob}

\vspace{0.5ex}
\vspace{-1ex}

Based on Eq. (\ref{eq:ES_AFDM_Rx_AD}), we can get

\vspace{-3ex}
 
\begin{eqnarray}
{H}_{i}^0[p,q]=\frac1Ne^{j\frac{2\pi}N\left(Nc_1(l^B_i)^2-ql^B_i\right)}{\mathcal{F}}_{i,B}[p,q],
\end{eqnarray} 
and 
\vspace{-3ex}

\begin{eqnarray}
\begin{aligned}
\begin{split}
&{{{{{H}}}}_{i,B}}[p,q] ={ H}_i^0[p,q]{e^{j2\pi \left[ {{{{c}}_2^A}[q]{q^2} - {{{c}}_2^B}[p]{p^2}} \right]}}\\
&=\frac1N {e^{j2\pi \left[ {{{{c}}_2^A}[q]]{q^2} - {{{c}}_2^B}[p]{p^2}} \right]}}{e^{j\frac{{2\pi }}{N}\left( {N{c_1}(l^B_i)^2 - q{l^B_i}} \right)}}{{\mathcal{F}}_{i,B}}[p,q],
\end{split}
\end{aligned}  
\end{eqnarray} 
where ${{\mathcal{F}}_{i,B}}\left[ {p,q} \right] {=} \frac{{{e^{ - j2\pi \left( {p - q - {\nu^B_i} + 2N{c_1}{l^B_i}} \right)}} - 1}}{{{e^{ - j\frac{{2\pi }}{N}\left( {p - q - {\nu^B_i} + 2N{c_1}{l^B_i}} \right)}} - 1}}$ with ${\nu ^B_i} = N{f^B_i}$. Thus, the input-output relation can be expressed as (\ref{eq:in_out_rel_Bob}) on the next page. We can see from (\ref{eq:in_out_rel_Bob}) that the vector ${\mathbf{c}}_2^A$ generated at the Alice affects every received symbol at the Bob.

\begin{figure*}[htbp]
	\begin{align} \label{eq:in_out_rel_Bob}
		{{y}}[p] &= \sum\limits_{i = 1}^P {{h^B_i}} \sum\limits_{q = 0}^{N - 1} {{{{H}}_i}} [p,q]{{x}}[q] + {{w}}[p] 
		= \sum\limits_{i = 1}^P {{h^B_i}{e^{-j2\pi {{{c}}_2^B}[p]{p^2}}}} \sum\limits_{q = 0}^{N - 1} {{e^{j2\pi {{{c}}_2^A}[q]{q^2}}}{e^{j\frac{{2\pi }}{N}\left( {N{c_1}(l^B_i)^2 - q{l^B_i}} \right)}}{{\mathcal{F}}_{i,B}}[p,q]} {{x}}[q] + {{w}}[p].
	\end{align}  
	\hrule
\end{figure*}

Since vectors ${\bf{c}}_2^A$ and ${\bf{c}}_2^B$ are controlled by two different LPPN generators, respectively, if these two LPPN generators are synchronized, we can get synchronized ${{\bf{c}}_2^A}$ and ${{\bf{c}}_2^B}$, i.e., ${{{c}}_2^A}[q] = {{{c}}_2^B}[q], q \in \left[0,N-1\right]$. As a result, the effective channel $\mathbf{H}_{{\rm eff},B}$ can be rewritten as

\vspace{-4ex}

\begin{align}
	&{H}_{{\rm eff},B} \left[p,q\right] = \sum\limits_{i = 1}^P {{{\tilde h}^B_i}  {H}_{i,B}\left[p,q\right]}  \\  
	&= \hspace{-1mm} \sum \limits_{i = 1}^P \hspace{-1mm} \frac1N  \hspace{-0.5mm}{{\tilde h}^B_i}\hspace{-0.5mm} {e^{j2\pi \left[ {{{{c}}_2^B}[q]{q^2} \hspace{-0.3mm} - \hspace{-0.3mm} {{{c}}_2^B}[p]{p^2}} \right]}}\hspace{-1mm}\times\hspace{-1mm} {e^{j\frac{{2\pi }}{N}\left( {N{c_1}(l^B_i)^2 \hspace{-0.3mm}-\hspace{-0.3mm} q{l^B_i}} \right)}}\hspace{-1mm}{\mathcal{F}_{i,B}}[p,q].\nonumber
\end{align}




Now, the effective channel is only affected by the vector ${{\bf{c}}_2^B}$ at the Bob. Hence, Bob can detect the $\mathbf{x}$ in the minimum mean square error (MMSE) criterion as follows
\begin{eqnarray}
{\hat {\bf{x}}}_B  = \mathbf{H}_{{\rm eff},B}^H{\left( {{\mathbf{H}_{{\rm eff},B}}\mathbf{H}_{{\rm eff},B}^H + \sigma _{n,B}^2{{\bf{I}}_N}} \right)^{ - 1}}{{\bf{y}}_B}.
\end{eqnarray}

It is shown that Bob can recover the transmitted symbols $\bf{x}$ from the received $\bf{y}_{B}$, after two LPPN generators of the Alice and the Bob are synchronized. The synchronization methods of LPPN have been widely studied \cite{raju2012digital}. Thus, we will discuss the synchronization methods in our future work due to page limits.

\vspace{-1ex}

\section{Security Analyses of SE-AFDM System}

In this section, we analyze the security of the proposed SE-AFDM system. Specifically, the input-output relation of SE-AFDM between Alice and Eve is derived. Based on this, we reveal that the effect of the vector ${\bf c}^A_2$ can not be eliminated at the Eve. Moreover, the effective SINR of Eve is analyzed, which shows the security is enhanced by reducing the eavesdropping quality of Eve.


\subsection{Input-Output Relation of SE-AFDM at the Eve}

{\bf Assumption 1:} It is assumed that Eve has a very strong capability, that is, it knows the fixed waveform parameters, e.g., $c_1, N$ and the codebook $\mathbb{C}_2$. However, Eve does not know the detailed structure of the LPPN generator of Alice, which means that Eve is unable to reconstruct and synchronize ${\bf c}^A_2$.


The received SE-AFDM signal at the Eve in the time domain can be expressed as 
\begin{equation} 
{\bf{r}}_{E} = \sum\limits_{i = 1}^P {{{\tilde h}_i^E}{{\bf{\Gamma }}_{{\rm cpp}_i}}{{\bf{\Delta }}_{{f_{i}^E}}}{{\bf{\Pi }}^{{l_i^E}}}{\bf{s}}_{A}}  + {\bf{w}}_{E} ,
\label{eq: relation between R and x in matrix form_Eve}
\end{equation}
where $\mathbf{w}_{E}\in {{\mathbb{C}}^{N\times 1}}$ is an additive Gaussian noise vector with power spectral density $\sigma _{n,E}^2$.  


Similar to (\ref{eq:ES_AFDM_Rx_TD}), after serial to parallel conversion (S/P) and discarding CPP, multiplying by matrix ${\bm{\Lambda} _{{c_1}}}$ and performing DFT, we can get
\begin{equation} \label{eq:ES_AFDM_Eve}
	{\bf{r}}'_{E}= \sum\limits_{i = 1}^P {{{\tilde h}_i^E} {\bf F}{\bm{\Lambda} _{{c_1}}} {{\bf{\Gamma }}_{{\rm cpp}{_i}}}{{\bf{\Delta }}_{{f^E_{i}}}}{{\bf{\Pi }}^{{l^E_i}}}{{\bf \Lambda} ^{H}_{{c_1}}} {\bf F}^{ H}{{\bf \Lambda} ^{ H}_{{c_2,A}}}{\bf x} }  + {\bf{w}}'_{E},
\end{equation}
where ${\bf{w}}'_{E} = {\bf F}{{\bf \Lambda} _{{c_1}}} {\bf{w}}_{E}$.

Then, ${\bf{r}}'_{E}$ is multiplied by matrix ${\bm{\Lambda} _{{c_2,E}}} \hspace{-0.2mm} {=} \hspace{-0.2mm} {\rm diag} \hspace{-1mm}\left( \hspace{-1mm} {{e^{ - j2\pi{c}^E_2\left[m\right]{m^2}}}\hspace{-1mm} ,\hspace{-0.5mm} m\hspace{-1mm} =\hspace{-1mm} 0, \hspace{-0.5mm} \ldots , \hspace{-0.5mm} N {-} 1} \hspace{-1mm}\right)$, where ${\bf c}^E_2$ is the parameter vector at the Eve, the received matrix at the Evea in the affine domain can be written in matrix form as

\vspace{-1ex}

\begin{align} \label{eq:ES_AFDM_Rx_EVE}
{\bf y}_{E}&= \sum\limits_{i = 1}^P {{{\tilde h}^E_i}{{\bf \Lambda} _{{c_2,E}}} {\bf F}{\bm{\Lambda} _{{c_1}}} {{\bf{\Gamma }}_{{\rm cpp}{_i}}}{{\bf{\Delta }}_{{f^E_{i}}}}{{\bf{\Pi }}^{{l^E_i}}}{\bm{\Lambda} ^{ H}_{{c_1}}} {\bf F}^{ H}{\bm{\Lambda} ^{ H}_{{c_2,A}}}{\bf x} }  + {\bf{\bar w}}_{E} \nonumber \\
& = {\bf H}_{{\rm eff},E}' {\bf x}'  + {\bf{\bar w}}_{E},
\end{align}
where ${\bf H}_{{\rm eff},E}' \hspace{-0.5ex}= \hspace{-0.5ex}\sum\limits_{i = 1}^P {{{\tilde h}^E_i}{\bm{\Lambda} _{{c_2,E}}} {\bf F}{\bm{\Lambda} _{{c_1}}} {{\bf{\Gamma }}_{{\rm cpp}{_i}}}{{\bf{\Delta }}_{{f^E_{i}}}}{{\bf{\Pi }}^{{l^E_i}}}\hspace{-0.5ex}{\bm{\Lambda} ^{ H}_{{c_1}}} {\bf F}^{H}} $,\hspace{-0.1ex} ${{\bf{x}}' }\hspace{-0.2ex} =\hspace{-0.2ex} {\bf{x}} \hspace{-0.1ex}\odot \hspace{-0.1ex}\bm{x}_c $, ${x}_c[q] = {e^{j2\pi {{{c}}_2^A}[q]{q^2}}}, q \in \left[0,N-1\right]$, and ${\bf{\bar w}}_{E} = {\bm{\Lambda} _{{c_2,E}}}{\bf F}{\bm{\Lambda} _{{c_1}}} {\bf{w}}_{E}$.
 

\subsection{Analyzing of the Effect of ${\bf c}^A_2$ on Eve}



According to Assumption 1, Eve can know the matrix ${\bf H}_{{\rm eff},E}'$. Hence, the vector ${{\bf{x}}^\prime }$ can be estimated by Eve using MMSE method as
\begin{eqnarray}
{\bf{\hat x}}_E^\prime  = {\bf H}_{{\rm eff},E}'^H{\left( {{{\bf H}_{{\rm eff},E}'}{\bf H}_{{\rm eff},E}'^H + \sigma _{n,E}^2{{\bf{I}}_N}} \right)^{ - 1}}{{{\bf y}}_E}.
\end{eqnarray}
Consequently, it can get
\begin{eqnarray}\label{eq:c2_x}
\left\{\begin{array}{rl}
e^{j2\pi {c}^A_2\left[0\right]0^2} \cdot x[0]  &=\hat{{x}}_E^{'}[0]\\
&\vdots\\
e^{j2\pi {c}^A_2\left[N-1\right](N-1)^2} \cdot  x[N-1] &=\hat{{x}}_E^{'}[N-1].
\end{array} \right.
\end{eqnarray}
We can see from (\ref{eq:c2_x}) that the received $\hat{{x}}_E^{'}[q]$ consists of both transmitted information symbol ${x}\left[q\right]$ and ${c}^A_2\left[q\right]$ for $q \in \left[1,N-1\right]$. If both ${\bf x}$ and ${\bf c}^A_2$ are varying and unknown to the Eve, the Eve cannot recover ${\bf x}$ and ${\bf c}^A_2$ according to received $\hat{\mathbf{x}}_E^{'}$, since each equation has two unknowns when $q\geq1$.



\subsection{Analysis of Effective SINR of Eve}


This paper reveals the security of SE-AFDM system by analyzing the effective SINR degradation of Eve. The effective SINR at the receiver is an important measure to characterize the system performance \cite{zou2007compensation}. When the effective SINR of Eve decreases, less information is eavesdropped due to the reduced BER at Eve. The effective SINR analysis is based on the additive white Gaussian noise (AWGN) channel \footnote{The characteristics obtained in AWGN channel is still consistent with our simulation results in the multipath fading channel.}. The analysis of the effective SINR of Eve in multi-path fading channel will be presented in our future work.

The output SINR at the Bob can be expressed as
\begin{eqnarray}\label{eq:SINR_B}
\mathrm{SINR}_B = \frac{{{p_s} \alpha ^2_B }}{{\sigma _{n,B}^2}} ={\gamma} _B,
\end{eqnarray}
where ${p_s}$ is the transmit power of Alice, $\alpha_B$ represents the large-scale fading from Alice to Bob, and $\gamma_B = {p_s} \alpha ^2_B/{{\sigma _{n,B}^2}}$ denoting the output signal-to-noise ratio (SNR) of the received signal at Bob.


At the Eve, the estimation of the $p$-th symbol can be rewritten as
\begin{align}
{\hat{x}}_E^{\prime}\left[p\right]& =  {{x}}_E^{\prime}\left[p\right] + w_E\left[p\right] = {x}[p]e^{j2\pi{c}^A_2[p]p^2}+w_E\left[p\right] \nonumber\\
&={x}[p]+(e^{j2\pi{c}^A_2[p]p^2}-1){x}\left[p\right]+w_E\left[p\right],
\end{align}
where $w_E[p]$ denotes the residual noise after symbol detection.

Therefore, the effective output SINR of the $p$-th symbol at Eve after signal processing is given by \cite[Eq. (16)]{zou2007compensation}
\begin{align}
	\label{Eq25}
\mathrm{SINR}_{E,p}& =\frac{{\mathbb{E} \left\{ \left| {x}\left[p\right] \right|^2 \right\} }}{{\mathbb{E} \left\{ \left| {x}\left[p\right] \right|^2 \right\} } \mathbb{E} \left\{\mid e^{j2\pi{c}^A_2[p]p^2}-1\mid^2\right\}+{\sigma_{n,E}^2}}.
\end{align}\normalsize

Through theoretical derivation, Eq.[25] is reformulated as

\begin{align}\label{eq:SINR_E_local}
\mathrm{SINR}_{E,p} \stackrel{a}=\frac{{p_s\alpha ^2_E}}{{2p_s\alpha ^2_E\left(1-\mathrm{Sa}\left(2\pi p^2c_{2,{\rm max}}\right)\right)}+{\sigma_{n,E}^{2}}},
\end{align}\normalsize
where $\alpha_E$ represents the large-scale fading from Alice to Eve, $\mathrm{Sa\left(x\right) = \frac{sin\left(x\right)}{x}}$ denoting the $\mathrm{Sa}$ function, 
and the equality of (a) holds when ${c}^A_2(p)$ is uniformly distributed over [$-c_{2,{\rm max}}$, $c_{2,{\rm max}}$]. 

The average effective SINR at the Eve is expressed as      
\begin{align}\label{eq:SINR_E}
\mathrm{SINR}_{E} = \frac{1}{N} \sum\limits_{p = 0}^{N-1} \frac{{\gamma _E}}{{2\gamma _E\left(1-\mathrm{Sa}\left(2\pi p^2c_{2,{\rm max}}\right)\right)}+1},
\end{align}
where $\gamma_E = {p_s} \alpha ^2_E/{{\sigma _{n,E}^2}}$  denotes the SNR of the received signal at Eve.


{\bf Discussion about effective SINR of Eve:} We can see from (\ref{eq:SINR_E_local}) and (\ref{eq:SINR_E}) that the effective SINR of Eve in SE-AFDM system is affected by $c_{2,{\rm max}}$, according to the property of the $\mathrm{Sa}$ function.

(\romannumeral1) When $c_{2,{\rm max}}$ is very small and tends to zero, i.e., $c_{2,{\rm max}} \to 0$, $\mathrm{Sa}\left(2\pi p^2c_{2,{\rm max}}\right) = 1$, and then $\mathrm{SINR}_{E} = \gamma _E$ = ${p_s} \alpha ^2_E/{{\sigma _{n,E}^2}}$. When the transmission power ${p_s}$ at Alice increases, the SNR of Bob $\gamma_B$ and the SNR of Eve $\gamma_E$ rise. At this time, there is a high risk of eavesdropping, which means that there is no security.


(\romannumeral2) Except for $p = 0$, when $c_{2,{\rm max}}$ is large enough, $\mathrm{Sa}\left(2\pi p^2c_{2,{\rm max}}\right) \to 0$, and then $\mathrm{SINR}_{E,p} = \frac{\gamma_E}{2\gamma_E + 1}$ . Furthermore, when $\gamma_E = {p_s} \alpha ^2_E/{{\sigma _{n,E}^2}} \gg 1$, $\mathrm{SINR}_{E,p} \approx 0.5$. At this time, as the transmission power ${p_s}$  at Alice increases, the SNR of Bob $\gamma_B$ grows while the SNR of Eve $\gamma_E$ remains relatively stable. With the amplified ${p_s}$, the relative stability of $\gamma_E$ enhances communication security by preventing eavesdropping quality from rapidly increasing.



In summary, only a large enough $c_{2,{\rm max}}$ can provide security in SE-AFDM system. Numerical results will verify this conclusion.

Next, we briefly analyze the complexity of brute force at the Eve and the spectrum efficiency of our SE-AFDM system.
If Eve searches for ${\bf c}^A_2$ in brute force method, the search space size is $M^N$ for the proposed SE-AFDM system. As a comparison, the search space size for a direct sequence spread spectrum (DSSS) system using $N$-length LPPN sequence is $2^N$. Moreover, our SE-AFDM system enjoys the same spectrum efficiency as the existing AFDM system.

\section{Simulation Results}
In this section, numerical results based on Monte Carlo simulations are presented. In our simulation, the carrier frequency $f_c = 24$ GHz, and the subcarrier spacing $\Delta _f = 15$ kHz. The QPSK symbols are transmitted. Unless otherwise specified, we consider a channel with $P=3$ paths, whose delays $l=\left[0,1,2\right]$ and maximum integer part of the normalized Doppler shift is $\alpha _{\rm max}$ = 2 corresponding to a maximum speed of 1350 km/h \cite{bemani2023affine}. Each path has a different Doppler shift generated by the Jakes model, i.e., $\nu _i = \alpha _{\rm max} cos(\theta _i)$, where $\theta _i$ is uniformly distributed over $\left[-\pi,\pi\right]$\cite{bemani2023affine}. The complex gain of the $i$-th path ${{h}_{i}}$ is set to be independent complex Gaussian random variables with zero mean and $1/P$ variance. We compare performances of our SE-AFDM  and the existing AFDM \cite{bemani2023affine}.

\begin{figure}[H]
	\vspace{-5pt}
	\centering
	\includegraphics[width=2.1in]{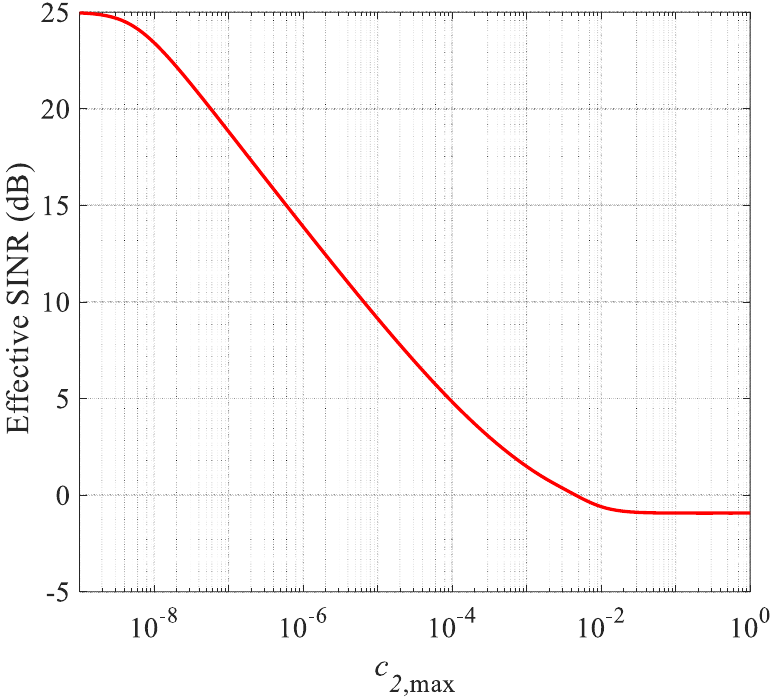}
	\caption{The effective SINR at Eve versus $c_{2,\rm max}$ of SE-AFDM with $\gamma _E$ = 25 dB.
		\label{fg:cs_vs_c2max}}
	\vspace{-5pt}
\end{figure}

Firstly, based on Eq. (\ref{eq:SINR_E}), the effective SINR at Eve in the SE-AFDM system with different $c_{2,\rm max}$ is shown in Fig. 2, where $\gamma _E$ $= 25$ dB and $P=1$ . It shows that the effective SINR of Eve declines with the increase of $c_{2,{\rm max}}$. 
 Consistent with the theoretical analysis, the effective SINR eventually approaches -0.93 dB as $c_{2,\rm max}$ continues to increase. 

\begin{figure}[H]
		\centering
		\vspace{-5pt}
	\includegraphics[width=2.1in]{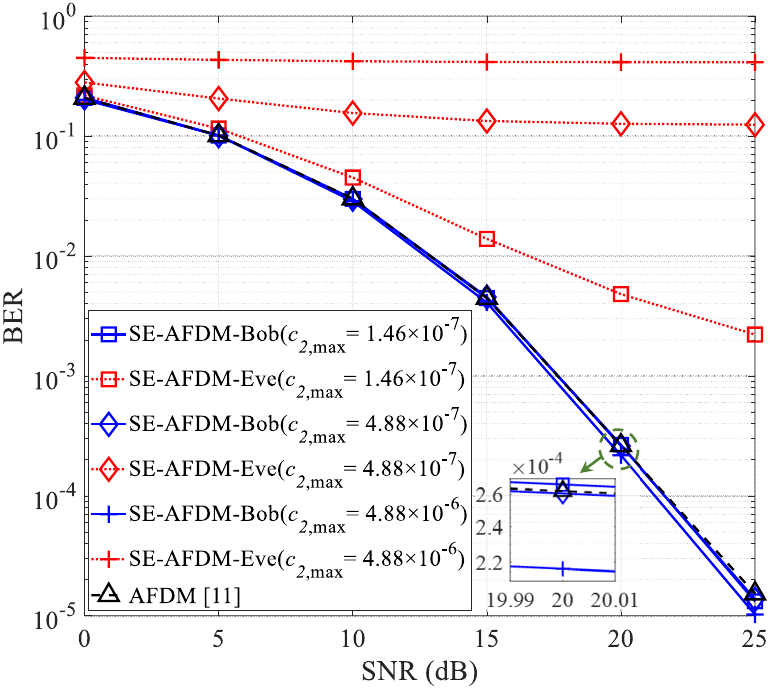}
	\caption{The BER performances versus SNR with different $c_{2,\rm max}$ of our SE-AFDM and the existing AFDM.
		\label{fg:BER_vs_SNR_diff_c2}}
	\vspace{-5pt}
\end{figure}

%

%

%


The BER performances versus SNR with different $c_{2,\rm max}$ are shown in Fig. \ref{fg:BER_vs_SNR_diff_c2}. In our proposed SE-AFDM system, the BER performances at the Bob are almost the same as that of the existing AFDM system for any $c_{2,\rm max}$. However, as $c_{2,\rm max}$ increases, the BER performance at the Eve deteriorates severely, tending to 0.5. These BER performances shows the security performance of the SE-AFDM system improves with the growth of $c_{2,\rm max}$, which is consistent with Fig. \ref{fg:cs_vs_c2max}.



\begin{figure}[h]
	\centering
	\includegraphics[width=2.1in]{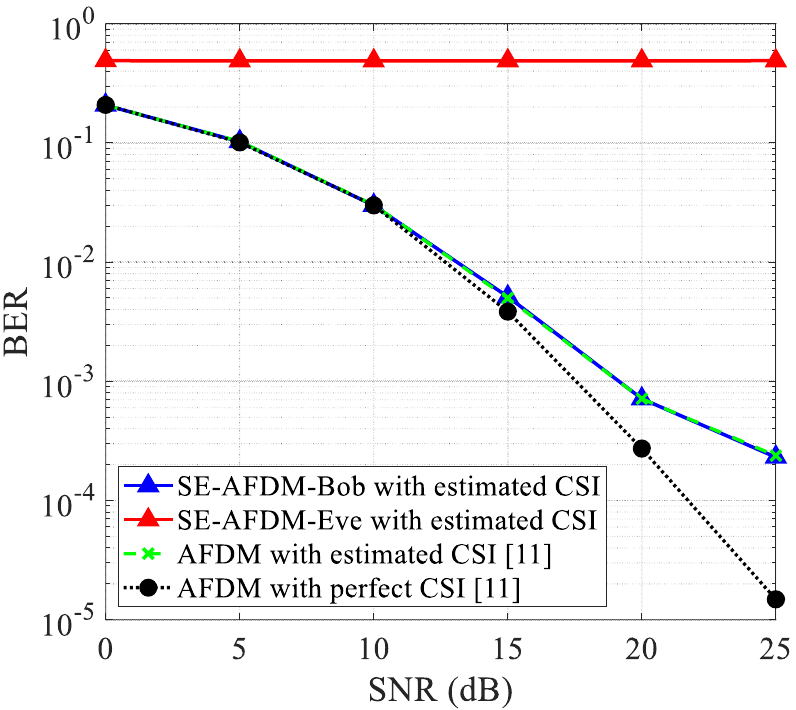}
	\caption{The BER performances of SE-AFDM and AFDM with estimated CSI at SNR$\rm_p$ = 30dB.
			\label{fg:var_doppler}}
			\vspace{-18pt}
	\end{figure}


Next, the impact of channel estimation errors on the BER performance of the SE-AFDM system is explored in Fig. \ref{fg:var_doppler} by adopting the channel estimation method \cite{bemani2023affine}. In this case, $c_{2,\rm max}$ is set as $4.88\times 10^{-6}$ and the SNR of pilot symbol, i.e., SNR$\rm_p$, is 30 dB.
With estimated CSI, the BER of our SE-AFDM system at the Bob coincides with that of the AFDM system in \cite{bemani2023affine}, both of which are slightly worse than the ideal BER performance. And the BER of the SE-AFDM system at the Eve still tends to be 0.5 with estimated CSI. It is shown that our SE-AFDM system still works in practically estimated channel scenarios.

 \vspace{-2pt}
 
 



\begin{figure}[htbp]
	\centering
	\includegraphics[width=2.1in]{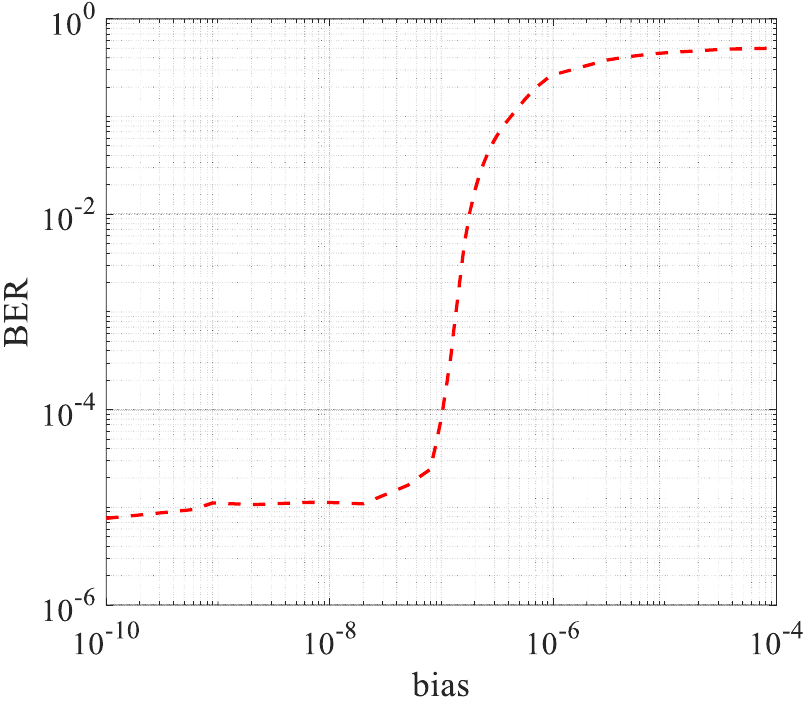}
	\caption{The BER performance of the Eve versus the bias $\sigma _{c_2}$ with SNR = 25 dB.
			\label{fg:bias}}
			\vspace{-9pt}
	\end{figure}

Finally, we investigate the impact of the bias of ${\bf c}^A_2$ on BER performance at the Eve, when Eve uses brute force method to search for ${\bf c}^A_2$. 
Here, $c_{2,\rm max}$ = $4.88\times 10^{-5}$, and the bias denotes the difference between actual ${\bf c}^A_2$ at the Alice and the ${\bf c}^E_2$ searched  by Eve, which is defined as $\sigma _{c_2} = \| { \bf c}^A_2 - {\bf c}^E_2 \|_{\infty}$. 
We can see from Fig. \ref{fg:bias} that the BER at the Eve is larger than 0.1 when the bias $\sigma _{c_2}$ is larger than $4.2\times10^{-7}$. As the bias decreases, the BER at the Eve reduces. When the bias is less than $2\times10^{-8}$, the BER remains constant and is close to the BER at the Bob. This result is helpful for designing the codebook $\mathbb{C}_2$, that is, 
enhanced security can be achieved by setting the interval between the alternative $c_2$ in the codebook $\mathbb{C}_2$ larger than a threshold, e.g., $4.2\times10^{-7}$.


\section{Conclusion}

This paper introduced a SE-AFDM communication system to improve the security of communication by using the time-varying parameter $c_2$. Our SE-AFDM system could significantly improve communication security by configuring appropriate parameter $c_2$. Numerical results showed that there is no lose in BER performance at the Bob compared with the existing AFDM system, but Eve can not eavesdrop on information from Alice.


\section*{Acknowledgment}

This work was supported in part by the China Postdoctoral Science Foundation under Grant Number 2024M764088, in part by the National Natural Science Foundation of China under Grant 61971025.



%

%
%

\bibliographystyle{IEEEtran}
\bibliography{IEEEabrv,refer}

\end{document}